\documentclass[sigconf]{acmart}

\usepackage{todonotes}
\usepackage{amsmath}
\usepackage{caption}
\usepackage{subcaption}
\usepackage{float}
\usepackage[inline]{enumitem}

\setcopyright{none}

\AtBeginDocument{%
  \providecommand\BibTeX{{%
    \normalfont B\kern-0.5em{\scshape i\kern-0.25em b}\kern-0.8em\TeX}}}

\acmYear{}
\acmDOI{}

\acmConference[KDD Workshop on Machine Learning in Finance 2021]{KDD Workshop on Machine Learning in Finance}{August, 2021}{Virtual}
\acmPrice{}
\acmISBN{}

\begin{document}

\title[Discovering material information on financial regulatory filings]{Discovering material information using hierarchical Reformer model on financial regulatory filings}

\author{François Mercier}
\affiliation{%
  \institution{Autorité des Marchés Financiers / Mila}
%  \streetaddress{P.O. Box 1212}
  \state{Québec}
  \country{Canada}
}
\email{francois.mercier@lautorite.qc.ca}

\author{Makesh Narsimhan}
\affiliation{%
  \institution{Autorité des Marchés Financiers / Mila}
%  \streetaddress{P.O. Box 1212}
  \state{Québec}
  \country{Canada}
}
\email{MakeshSreedhar.Narsimhan@lautorite.qc.ca}

\renewcommand{\shortauthors}{Mercier and Narsimhan}

\begin{abstract}
    Most applications of machine learning for finance are related to forecasting tasks for investment decisions. Instead, we aim to promote a better understanding of financial markets with machine learning techniques. Leveraging the tremendous progress in deep learning models for natural language processing, we construct a hierarchical Reformer (\cite{kitaev2020reformer}) model capable of processing a large document level dataset, SEDAR,  from canadian financial regulatory filings. Using this model, we show that it is possible to predict trade volume changes using regulatory filings. We adapt the pretraining task of HiBERT (\citep{hibert}) to obtain good sentence level representations using a large unlabelled document dataset. Finetuning the model to successfully predict trade volume changes indicates that the model captures a view from financial markets and processing regulatory filings is beneficial. Analyzing the attention patterns of our model reveals that it is able to detect some indications of material information without explicit training, which is highly relevant for investors and also for the market surveillance mandate of financial regulators.
\end{abstract}



\keywords{NLP, Transfer Learning, Financial Reports, Self-Supervised Learning, Material Information detection}


\maketitle

\section{Introduction}
The entire financial services industry is based on the trust investors have in the system. Preserving this trust is a core mandate from all regulators across the world. Machine learning techniques have the potential to improve several aspects in this industry, including trust. However, most interest so far has been on forecasting tasks for investment decisions.

 A better understanding of the markets is useful for everyone in the system. For investors, this translates to a more accurate analysis of the risk-return trade-off. For regulators, it helps to focus their efforts on  detecting potential market manipulations and ensuring trust in the overall market. 

In this work, we focus on providing answers to the following research questions:
\begin{enumerate}
    \item \textit{Can we predict market events using only a large document-level dataset from regulatory filings ?}

    \item \textit{If yes, can we leverage potential predictive power to get insights from markets about filings?}

\end{enumerate}

The main contributions of our work include:
\begin{itemize}
    \item Using a hierarchical model, inspired by Hibert (\citet{DBLP:journals/corr/abs-1905-06566}) and leveraging the efficient Reformer model (\citet{kitaev2020reformer}) for obtaining good representations long sequences and using these obtained representations successfully for a surrogate downstream task, predicting direction of trade volume changes from potentially long documents
    \item Providing a qualitative assessment of predictions and using attention patterns to better understand the market point of view on focused documents and sentences.
\end{itemize}

\section{Problem definition}

\subsection{Background}

"Material change", according to securities legislation, is defined as “a change in the business, operations or capital of the issuer that would reasonably be expected to have a significant effect on the market price or value of any of the securities of the issuer and includes a decision to implement such a change made by the board of directors of the issuer by senior management of the issuer who believe that confirmation of the decision by the board of directors is probable”. Regulators require issuers to disclose immediately these "material changes" in order to ensure a level playing field for all investors, and therefore, to ensure trust in financial markets.

% https://lautorite.qc.ca/fileadmin/lautorite/reglementation/valeurs-mobilieres/51-201/2013-05-31/2013mai31-51-201-ig-vadmin-en.pdf
\subsection{Formal definition}

In order to better understand potential material changes, we define the following surrogate downstream task: using documents, publicly available for investors, we try to predict the trade volume movement within a 1 business day time horizon from the release date $D$ for the stock associated with this document. Let $V_{D}$ the daily volume for a specific stock for the day $D$.

\begin{align}
    &\textit{daily volume change} = V_{D+1} - V_{D} \nonumber\\
    &Y_D = 
    \begin{cases}
    1 \quad \textit{if daily volume change} \ge 0\\
    0 \quad \textit{otherwise}
    \end{cases}  \label{equation:1}
\end{align}
The model for this surrogate downstream task is designed to extract focus information at sentences level (see section \ref{section:model}). The sentences receiving the most focus are then selected to propose an extractive summary of these documents, ideally with potential further information about material information.
\section{Related work}

\begin{figure*}[h]
  \centering
\centerline{\includegraphics[width=.8\textwidth]{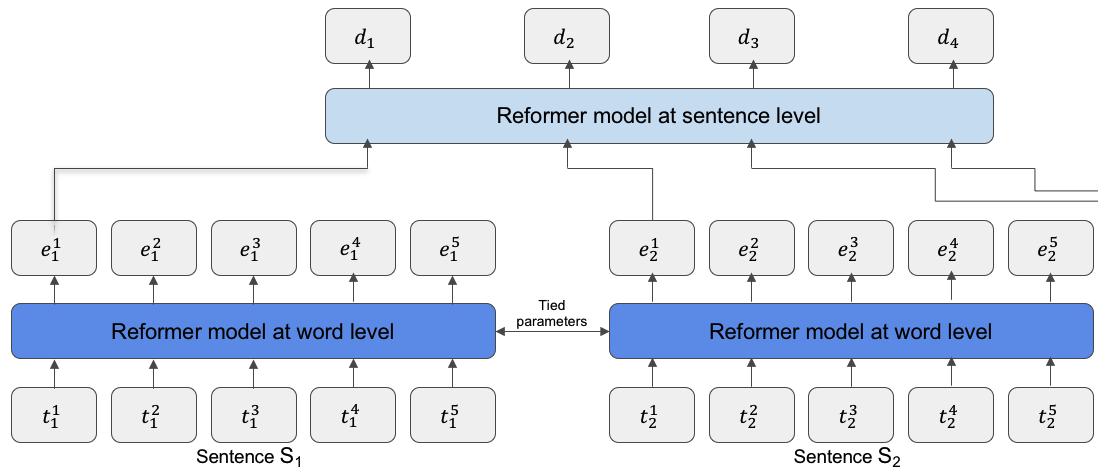}}
  \caption{Base hierarchical model to obtain contextualized sentences embeddings.} \label{image:base-model}
\end{figure*}

\subsection{Market signals and market efficiency}

Finance is the management of money by companies, organisations, or governments, as defined by Wikipedia\footnote{\href{https://en.wikipedia.org/wiki/Finance}{https://en.wikipedia.org/wiki/Finance}}. It relates to spending and investment of capital, which are decisions made from opinions. Decision processes may be the result of human opinions directly, with manual trade executions, or indirectly, thanks to algorithmic executions using human prior knowledge. With this insight, from a market perspective, indicators, resulting from market activities, are the equilibrium from different investor opinions. These indicators are commonly referred to as market signals. Market signals can be trade prices, prices volatility, limit order book information,  or trade volumes.

The main challenge with using market signals, which can be used as direct inputs for investment decisions, is related to the Efficient Market Hypothesis (EMH), \citet{malkiel1970efficient}.  The EMH suggests that no one can beat the market without additional risk, which is described with the "No Free Lunch" principle. 
While there is still a debate regarding the validity of this hypothesis, \citet{doi:10.1111/j.1540-6261.1991.tb04636.x-fama1991,RePEc:eee:jfinec:v:49:y:1998:i:3:p:283-306,10.1257/089533003321164958}, research suggests that the semi strong form of EMH, where all past and present public information is reflected in market prices, seems to hold for developed markets, at least to some extent, see \citet{rossi2015efficient} for literature review. Thus, the consensus is that EMH appears to be a good approximation of market behaviours and one of the main reasons for the difficulty in discovering and using market signals.

In order to tackle this issue, several strategies are often used like incorporating domain knowledge, such as feature engineering with creation of domain specific lexicon from 10-K filings, \citet{doi:10.1111/j.1540-6261.2010.01625.x}, capturing specific events with open information extraction, \citet{ding2014using}, and also exploiting new dataset, referred as alternative data, such as Twitter to capture investors sentiment, \citet{10.1007/978-3-642-39146-0_8}. 

Due to the strong interest of exploiting market signals for investment decisions, most of research focused on price forecasting or price movement direction, (\citet{chong2017deep}, \cite{Jiang2020ApplicationsOD}, \citet{bartram2020artificial}). Some exceptions can be found for alternative market signals. For price volatility, \citet{theil-etal-2018-word} suggests SEC 10-k filings have got predictive power. \citet{6816733} suggests web traffic has predictive power for trade volumes. However, to the best of our knowledge, none are leveraging these market signals as interpretation from markets point of views.

\subsection{NLP}

 \paragraph{Contextual Word Embeddings} Recently, there has been a shift from using distributional word representations (\citet{word2vec,glove}), which result in a single global representation for each word ignoring their context, to contextual embeddings, where each token is associated with a representation that is a function of the entire input sequence. These context-dependent representations can capture many syntactic and semantic properties of words under diverse linguistic contexts. Previous work (\citet{Peters:2018,devlin-etal-2019-bert,joshi2020spanbert,lan2020albert,liu2019roberta,clark2020electra}) has shown that contextual embeddings pretrained on large-scale unlabelled corpora achieve state-of-the-art performance on a wide range of natural language processing tasks, such as text classification, question answering and text summarization. 
 \vspace{-4mm}
 \paragraph{Document level embeddings} 
   HIerachical Bidirectional Encoder Representations from Transformers (\citet{hibert}) builds upon BERT (\citet{devlin-etal-2019-bert}) and proposes a pretraining scheme for document level embeddings. To obtain the representation of a document, they use two encoders: a sentence encoder to transform each sentence in the document to a vector and a document encoder to learn sentence representations given their surrounding sentences as context. Both the sentence encoder and document encoder are based on the Transformer encoder (\citet{vaswani2017}) nested in a hierarchical fashion. They use a variant of the Masked Language Modeling paradigm using sentences as the basic unit instead of words. i.e. they predict masked out sentences given the context. They show that such a pretraining scheme is highly effective and allows them to achieve SOTA results on summarization tasks. 
    \vspace{-4mm}
  \paragraph{Self-Attention Variants} Recently, there has been a lot of interest in breaking the quadratic self attention used in transformer (\citet{kitaev2020reformer,beltagy2020longformer,wang2020linformer,zaheer2020big}) with lower time and memory complexities, enabling the processing of larger sequences and giving rise to better models.  Simplest methods in this category just employ a sliding window, but in general, most work fits into the following general paradigm: using some other mechanism select a smaller subset of relevant contexts to feed in the transformer and optionally iterate. 
In this work, we use the Reformer model which introduces the following improvements:
\begin{enumerate*}
     \item using reversible layers to remove the need to store intermediary activations for the backpropagation algorithm;
    \item splitting activations inside the feed-forward layers and processing them in chunks;
    \item approximating attention computation based on locality-sensitive hashing.
\end{enumerate*}

\section{Model} \label{section:model}
\subsection{Base hierarchical model: Document representation}

We use the following notation: $D = (S_1, S_2, ..., S_{|D|})$ for the sequence of sentences in a document,  $S_i = (t^1_i, t^2_i, ..., t^{|S_i|}_i)$ for the sequence of subword tokens for the $i$th sentence and $t^j_i$ for the $j$th subword token for the $i$th sentence.

The base model is the module shared by all the different tasks and is composed of two main submodules, both Reformer models (\citet{kitaev2020reformer}):
\begin{itemize}
    \item The first submodule transforms the sequence of subword embeddings $(t^1_i, t^2_i, ..., t^{|S_i|}_i)$ into a sequence of contextual embeddings at subword level, $(e^1_i, e^2_i, ..., e^{|S_i|}_i)$ for each sentence $S_i$. For later stages, the first embedding is treated as the sentence embedding for the $i$th sentence.
\begin{align*}
    (e^1_i, e^2_i, ..., e^{|S_i|}_i) = \textit{ReformerModel}_{word}(t^1_i, t^2_i, ..., t^{|S_i|}_i)
\end{align*}
\item The second submodule transforms a sequence of sentence embeddings $(e^1_1, e^1_2, ..., e^1_{|D|})$ into a sequence contextual embeddings at sentence level, $(d_1, d_2, ..., d_{|D|})$. 
\begin{align*}
    (d_1, d_2, ..., d_{|D|}) = \textit{ReformerModel}_{sentence}(e^1_1, e^1_2, ..., e^1_{|D|})
\end{align*}

\end{itemize}

To recap, the base model is defined as follow:
\begin{align*}
    (d_1, d_2, ..., d_{|D|}) = \textit{BaseModel} (D)
\end{align*}
For a graphical visualization, please see figure \ref{image:base-model}.

\subsection{Hierarchical model for the pretraining task}

\begin{figure}
\begin{minipage}[b]{0.48\textwidth}
     \centering
    \includegraphics[width=1\linewidth]{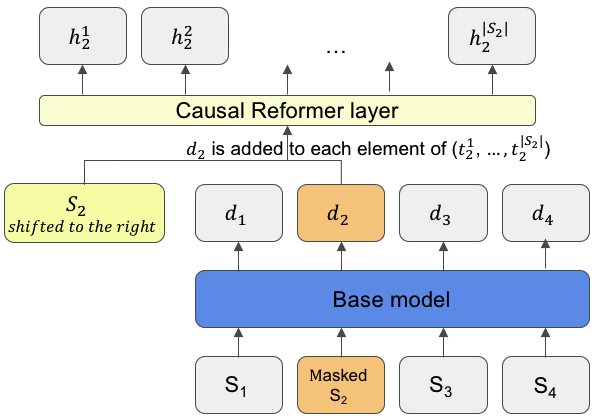}
      \caption{Hierarchical model for the pretraining task.} \label{image:hibert-model}
   \end{minipage} \hfill
   \begin{minipage}[b]{0.48\textwidth}
     \centering
        \includegraphics[width=0.8\linewidth]{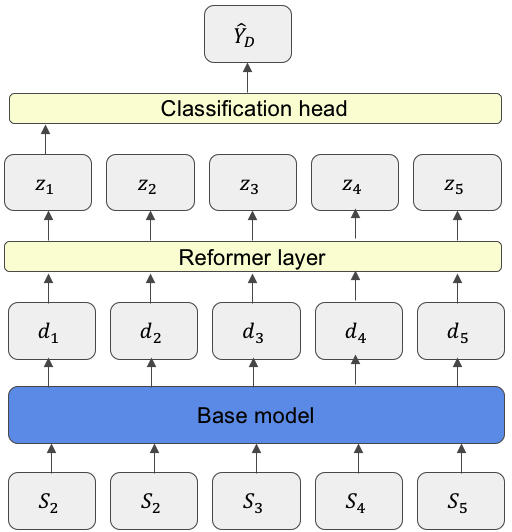}
          \caption{Hierarchical model for the surrogate downstream task (classification task).} \label{image:classification-model}
   \end{minipage}
\end{figure}

For the pretraining task, we used Hibert (\citet{DBLP:journals/corr/abs-1905-06566}) with the adaptation to long documents by using Reformer (\citet{kitaev2020reformer}) instead of Transformer (\citet{DBLP:journals/corr/VaswaniSPUJGKP17}).  

Let $M$ represents the set of indices of the masked sentences. 

For the masked sentences with indices $i \in M$, and for each subword tokens with indices $j \in [1; |S_i|]$, we replace all original tokens $t^j_i$  by the mask token \textit{<MASK>}. We then use a base model to get the contextual embeddings $\{d_i | i \in M\}$ at sentence level for these masked sentences. 

As the context, represented by $d_i$,  is a fixed vector for each subword prediction, this context is injected by adding this sentence embedding as well as $p^{j}$ position embedding for the $j$th indice for each timestep. We then use a single Reformer layer as follow:
\begin{align*}
    x_{i}^{j} &= \textit{EmbeddingLayer}(t_i^j) + d_i + p^{j}\\
    q_{i}^{j-1} = x_{i}^{j-1},\quad K_{i}^{j-1} &= (x_{i}^{1}, x_{i}^{2}, ..., x_{i}^{j-1}),\quad V_{i}^{j-1} = (x_{i}^{1}, x_{i}^{2}, ..., x_{i}^{j-1}) \\
    h_{i}^{j-1} &= \textit{ReformerLayer} ( q_{i}^{j-1}, K_{i}^{j-1}, V_{i}^{j-1})
\end{align*}

From this, we predict the next word given the masked sentence embedding $d_i$ and the previous words $t_i^{1:j-1}$ $\textit{with } W_{ff} \in \mathbb{R}^{vocab\_size * hidden\_size} \textit{ and } b_{ff} \in \mathbb{R}$:
\begin{align*}
    p(t_i^j| t_i^{1:j-1}, d_i) &= \textit{softmax}(W_{ff} h_i^{j-1} + b_{ff}) 
\end{align*}

We then compute the cross entropy loss to be minimized for this task and discard the additional Reformer layer for downstream tasks.
\begin{align*}
    \textit{Loss}_{CrossEntropy}(D) = -\frac{1}{|M|} \sum_{i \in M} \sum_{j=1}^{|S_i|} log \left( p(t^j_i| t^{1:j-1}_i, d_i) \right)
\end{align*}
For a graphical visualization, please see figure \ref{image:hibert-model}.

\subsection{Hierarchical model for the surrogate downstream task} \label{section:model_downstream_task}

The classification task aims to incorporate knowledge from market signals into the hierarchical model. For this task, for a document $D$, the label $Y_D$ is is constructed as per equation \ref{equation:1}.

On top of the base model, we add a single Reformer layer with global attention mechanism to retrieve a latent representation of the sequence of sentences $(z_1, z_2, ..., z_{|D|})$ for the document $D$. The attention weights $\alpha_i$ used by this Reformer layer will be used later to retrieve the focus of the model.
\begin{align*}
    (z_1, z_2, ..., z_{|D|}) &= \textit{ReformerLayer} (\textit{BaseModel(D)}) 
\end{align*}
We then treat the first latent embedding $Z_1$ for the document $D$ as the document embedding and use a classification head. This head outputs the predicted probability of increase or decrease of daily volume, used later for the binary cross entropy loss:
\begin{align*}
    \hat{Y}_D = p(\textit{daily volume change} \ge 0 | D) = \textit{sigmoid} \left( \textit{ClassficationHead} (z_1) \right) \\
    \textit{Loss}_{BCE}(D) = -\left(Y_D * log(\hat{Y}_D )+ (1 - Y_D) * log(1 - \hat{Y}_D)\right)
\end{align*}
Finally, to encourage sparsity in the attention weights in additional Reformer layer for this task, in order to increase the focus on the most important sentences, we add a $l_1$ regularization term for the attention weights $\alpha_i$.
\begin{align*}
    \textit{Loss}_{classification}(D) = \textit{Loss}_{BCE}(D) + \sum_i |\alpha_i|
\end{align*}
For a graphical visualization, please see figure \ref{image:classification-model}.
\section{Experiments and results}
\subsection{Dataset}

\subsubsection{SEDAR}

The System for Electronic Document Analysis and Retrieval (SEDAR)\footnote{\url{https://www.sedar.com}} is a filing system developed for the Canadian Securities Administrators\footnote{\url{https://www.securities-administrators.ca}} (CSA), umbrella organization of provincial and territorial regulators in Canada, to: facilitate the electronic filing of securities information as required by CSA; allow for the public dissemination of Canadian securities information collected in the securities filing process; and provide electronic communication between electronic filers, agents and the CSA. It can be viewed as the Canadian equivalent of EDGAR, the filing system managed by the SEC\footnote{\url{https://www.sec.gov/}}.

For the purpose of this work, we collected 3.8M documents in English from 1997 to October 2018. These documents were originally in PDF and have been converted into raw text format. This conversion has resulted in a certain amount of noise into the raw text.

For the pretraining task, we used 2M documents randomly selected from the overall dataset. All selected documents were before 2018 to prevent information leakage.
% 1000 for valid set but this is not relevant enough

For the downstream task, we only used News releases and Management, Discussion and Analysis (MD\&A) documents for firms part of the TSX S\&P 60 index as of 30th July 2020. The selected document types are among the most known ones to broadcast financial information that are not part of standardised accounting metrics (e.g.: net income, cash flow from operations, ...). They are therefore highly susceptible to contain new material information. The training set contained 14,241 documents before 2018. The validation and test set are constituted by all documents in 2018, from which a random allocation has been performed to have 50\% for validation set and 50\% for test set. This results into 612 documents for validation set and  613 documents for test set. 89.5\% of documents contain less than 512 sentences and 94\% sentences contain less than 128 subwords.

\subsubsection{Market data} 
For the downstream task, we collected daily trade volumes from 2004 from the Bloomberg Terminal for all components from the TSX S\&P 60 Index. We then computed the daily change as per equation \ref{equation:1} and joined this dataset with SEDAR using filing dates and ticker codes. Labels are fairly imbalanced with 59\%, 63.6\% and 57.8\% up movements for train, validation and test sets respectively.

\subsection{Implementation details}

\subsubsection{Tokenization} 
We used BBPE (\citet{DBLP:journals/corr/SennrichHB15}) implementation from HuggingFace\footnote{\href{https://github.com/huggingface/tokenizers}{Tokenizers library: https://github.com/huggingface/tokenizers}} with a vocabulary size 8,000. Similar to common practices in natural language processing, for each sentence $S_i$, \textit{<BOS>} and \textit{<EOS>} are added to represent the beginning of sentence and the end of sentence.

\subsubsection{Base model} 
The base model used for following hyperparameters: 8 attention heads, 8 layers (4 for the Reformer model at word level and 4 for the Reformer model at sentence level), intermediary size 2048, maximum sentence length 128 and maximum number of sentences 512. The number of parameters for the base model was 67M. We used the Reformer implementation from HuggingFace\footnote{\href{https://github.com/huggingface/transformers}{Transformers library: https://github.com/huggingface/transformers}}.

\subsubsection{Pretraining} 
We trained our model with 2M documents for 1 epoch with a learning rate of $2\mathrm{e}{-4}$ with a linear learning rate schedule and with a batch size 32 using gradient accumulation. The duration of the training was 20 days using a single GPU (11GB).

\subsubsection{Downstream task} 
After hyperparameter search, we trained our best model for 2 epochs with a learning rate of $3\mathrm{e}{-6}$ for models with frozen base model encoder and $2\mathrm{e}{-5}$ otherwise, with a $l_1$ factor 0.1, with a cosine annealing learning rate schedule and with a batch size 32 using gradient accumulation. We used 2 layers MLP for classification head. Each training run lasted less than 1 day using a single GPU (11GB).

\subsection{Evaluation methodology}

We evaluated our model performance on the downstream task using the test set with ROC-AUC, MCC and F1 metrics. For a better results reporting, we used bootstrapping to obtain confidence intervals by repeating 100 times the evaluation on 300 observations. Classifications thresholds were calibrated to maximize MCC and F1 on validation set.

\subsection{Results and discussion}
\begin{table*}
\centering
\caption{Test set results for volume direction classification with 95\% CI. \textit{Majority baseline uses the majority class from training set. Random init.: random weights initialization (no pretraining).}}
\label{table:results}
\begin{tabular}{lccc} 
    \toprule
Model                                                        & ROC-AUC & MCC  & F1 \\ 
    \cmidrule(r){1-4}
Random baseline                                            & [49.9\%, 51.0\%]    & [- 0.1\%, 2.1\%] & [53.2\%, 54.5\%]     \\
Majority baseline                                    & [50.0\%, 50.0\%]    & [ 0.0\%, 0.0\%] & \textbf{[72.7\%, 73.4\%]}    \\
Ours - random init. & \textbf{[56.9\%, 57.8\%]}    & \textbf{[12.8\%, 14.5\%]} &  \textbf{[72.5\%, 73.2\%]}     \\
%Ours - frozen base model with random init.          & $[54.7\%; 58.0\%]$    & $[6.4\%; 12.3\%]$ & $[54.3\%; 57.2\%]$     \\
%Ours - frozen base model pretrained with 600K docs      & $[57.7\%; 60.6\%]$    & $[15.8\%; 21.4\%]$ & $[56.3\%; 59.3\%]$     \\
%\textbf{Ours-frozen+pretraining}        & $\mathbf{[58.4\%; 61.2\%]}$    & $\mathbf{[11.6\%; 16.3\%]}$ & $\mathbf{[56.1\%; 58.4\%]}$     \\
    \bottomrule
\end{tabular}
\end{table*}

\begin{table*}
\centering
\caption{Test set results for volume direction classification with 95\% CI. \textit{Random init.: random weights initialization (no pretraining). Frozen: frozen base model.}}
\label{table:results-pretraining}
\begin{tabular}{lccc} 
    \toprule
Model                                                        & ROC-AUC & MCC  & F1 \\ 
    \cmidrule(r){1-4}
Ours - random init. & [56.9\%, 57.8\%]    & \textbf{[12.8\%, 14.5\%]} & \textbf{[72.5\%, 73.2\%]}     \\
Ours - frozen+random init.          & [55.1\%, 56.2\%]    & [ 3.5\%, 5.6\%] & \textbf{[72.7\%, 73.4\%]}     \\
Ours - frozen+pretrained 600K docs      & \textbf{[57.5\%, 58.4\%]}    & \textbf{[13.6\%,  15.2\%]} & \textbf{[72.2\%, 73.0\%]}     \\
Ours - frozen+pretrained 2M docs        & \textbf{[58.0\%, 59.0\%]}    & [11.6\%, 13.1\%] & \textbf{[72.2\%, 73.0\%]}     \\
    \bottomrule
\end{tabular}
\end{table*}

\subsubsection{Predicting market signals from regulatory filings} 
Our results, summarized in the table \ref{table:results}, confirm the predictive power from only using regulatory filings for trade volume change prediction. 

Our model, without pretraining, outperforms significantly "simple" baselines with the exception of F1 score. F1 score doesn't take into account true negatives, which penalizes our model compared to the majority class prediction model. Despite the former, our model still achieves similar performance statistically. We only keep F1 score for comparison as this metric is commonly reported in related studies. 

As a conclusion, our results support the importance of information from regulatory reporting, at least for News releases and MD\&A, for investors in their investment decisions.

\subsubsection{Pretraining on document dataset} 
In this experiment, we compare pretrained models with frozen base model with randomly initialized model with frozen base model and a fully trainable model, also randomly initialized.

As indicated by the table \ref{table:results-pretraining}, the model pretrained with 2M documents outperforms the fully trainable model with statistical significance on ROC-AUC metric. For MCC metric, the model pretrained with 600K documents is statistically similar to the fully trainable model. For the F1 score, all models are statistically similar, making this metric less useful for ranking. Therefore, models ranking depends on the selected metric, but pretrained models outperform the fully trainable model on all metrics, except MCC for which results are statistically similar.

Moreover, we observed that the best performing model, in term of validation loss, on the pretraining task was the one pretrained on 600K documents. This model is also the best on the MCC metric. For other metrics, it still remains statistically similar to the one pretrained on 2M documents. Whereas the common knowledge assumes that further pretraining helps, see \citet{DBLP:journals/corr/abs-1907-11692}, we didn't find it in our case. As model capacity limited by the single GPU memory constraint, our models were smaller in number of parameters than \citet{DBLP:journals/corr/abs-1905-06566}. 
Thus, we believe bigger models may provide better results for this task.

Furthermore, an interesting observation is that, we noticed that pretrained models have less variance on training set and validation set losses during training than the one with frozen weight and random initialization. This suggests that initialization from pretraining does help to have a more robust training.

We conclude that the Hibert (\citet{DBLP:journals/corr/abs-1905-06566}) pretraining task is slightly beneficial for learning good representation from financial text corpus. 

\subsubsection{Qualitative analysis} 
For this experiment, we analyzed the best predictions on validation set from our model, meaning the most confident predictions for increase and decrease trade volumes. We reviewed the documents qualitatively and we also looked at the attention weights at the Reformer layer, on top of the base model, which is specialized for the downstream task, see section \ref{section:model_downstream_task}. We applied two attention patterns analysis, one using raw attention weights values similar to the study of BERT attention weights from \citet{clark-etal-2019-bert} and one using attention vector norms (\citet{kobayashi2020attention}). For both analyses, results were fairly consistent, even if some orders slightly changed.

Despite our use of $l_1$ regularizer to encourage sparsity, attention patterns are quite broad among attention heads. This indicates that there is no strong focus on a single sentence, but rather a broad focus on all sentences. While the focus appears to contain some noise, we noticed the model tends to focus more on: the second sentence of the document; sentences containing information about the firm (ex: ticker, firm name, web site url); statements related to disclosures of risks (ex: "\textit{Forward-looking statements in this document include, but are not limited to, statements relating to our financial performance objectives, vision and strategic goals, and include our President and Chief Executive Officers statements.}"); and also sometimes on mentions to non "Generally Accepted Accounting Principles" (GAAP) measures. While focus on firm name information is quite intuitive, the focus on information about non GAAP measures is quite surprising. Indeed, non GAAP measures have received some attentions from literature and regulators for decades as they may mislead investors in some cases, \citet{Bradshaw2000GaapVT}, \citet{entwistle2005voluntary}, \citet{marques2006sec}. The level of noise in the attention pattern remains an open question. Some reasons may be due to the model capacity, our assumption of using only one document to predict market event without further context, or also to the intrinsic evolving nature of markets.

At the document level, the most confident predictions of decrease of trade volumes appear to be on documents containing no clear new information: such as date confirmation for a result announcement; termination of already announced third party share repurchase program; or third party recognition related to "Environmental, Social and Corporate Governance" (ESG). All of these documents contained no accounting or new business development information. For the most confident predictions of increase of trade volumes, documents contain new accounting information, such as increase of net income or debt refinancing. See figures in appendix for examples. The previously mentioned patterns are consistent with the common knowledge for fundamental analysis.

To conclude, attention weights patterns appear to be too noisy to be used as a clear extractive summary for documents with our approach. At the document level, our model predictions are inline with accepted knowledge about the focus of the market, despite no prior knowledge about markets and no strong labels. Thus, without the need of labelling documents, we believe that our model can be relevant for finding documents containing potential material information, which could help investors for their risk management and also regulators for their market surveillance mandate. 

\section{Conclusion}

By leveraging recent progress in natural language processing to process document level large dataset for the financial context, we have proposed a new approach to attempt to discover material information. The key ideas are: efficient deep learning models for long sequences, such as Reformer (\citet{kitaev2020reformer}) in our work; a hierarchical model for capturing sentence level and document level contextualized embeddings; and a surrogate downstream task to align market signals, volume prediction in our work, with financial filings text dataset. We also show the benefits of the HiBERT (\citet{DBLP:journals/corr/abs-1905-06566}) pretraining task to improve the quality of sentence level embeddings by using a large unlabelled financial corpus. Finally, while attention patterns learnt by our model are still noisy, we were able to demonstrate the ability to discover material information without prior knowledge, which is relevant to regulators for their market surveillance mandate. With this work, we hope to encourage research between deep learning and finance communities as benefits could deserve all actors in the financial industry, including regulators, and ultimately users.

% \section*{Broader Impact}

% \paragraph{Importance of existing datasets in finance for machine learning}
% In the financial industry, datasets have been created for decades, mainly to increase market trust, and usually due to regulators actions. While regulatory reporting comes with costs on market participants, we believe benefits could be also given back to the financial industry by further exploiting these datasets, either for researchers or practitioners. In this work, we hope to encourage this direction to better understand financial markets for the benefits of all.

% \paragraph{Potential further use cases by leveraging document level datasets}
% Thanks to recent progresses on more efficient models, such as Reformer (\citet{kitaev2020reformer}) in our work, we confirm the ability to process real world dataset for long documents. This opens new possibilities on applying machine learning, especially deep learning techniques, on financial text datasets, which are more document level datasets than sentence level dataset.

% \paragraph{Democratization of deep learning techniques}
% By constraining this work with low resource computing (single GPU), we hope to encourage further research and applications by more actors than the ones with large computing infrastructures, usually from very limited number of countries. Diversity of actors is a prerequisite for diversity of ideas, which improves spreading benefits to all. This point is relevant for all industries and not just for the financial industry.

%\section*{References}

\small

\bibliographystyle{plainnat} % https://tex.stackexchange.com/questions/111790/natbib-in-text-citation-displays-author
\bibliography{refs} % Entries are in the "refs.bib" file

%\clearpage

\onecolumn

\section*{Appendix}
\thispagestyle{empty}
\begin{figure}[H]
        \begin{subfigure}{1\linewidth}
        \includegraphics[width=1\textwidth]{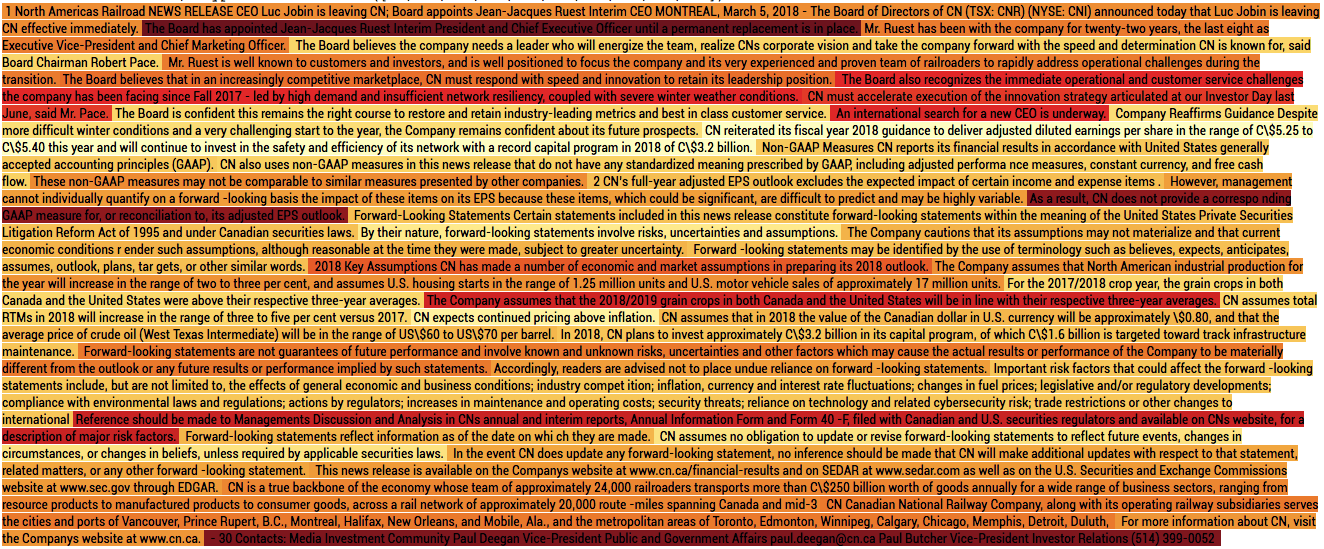}
    \end{subfigure}
        \begin{subfigure}{1\linewidth}
        \includegraphics[width=1\textwidth]{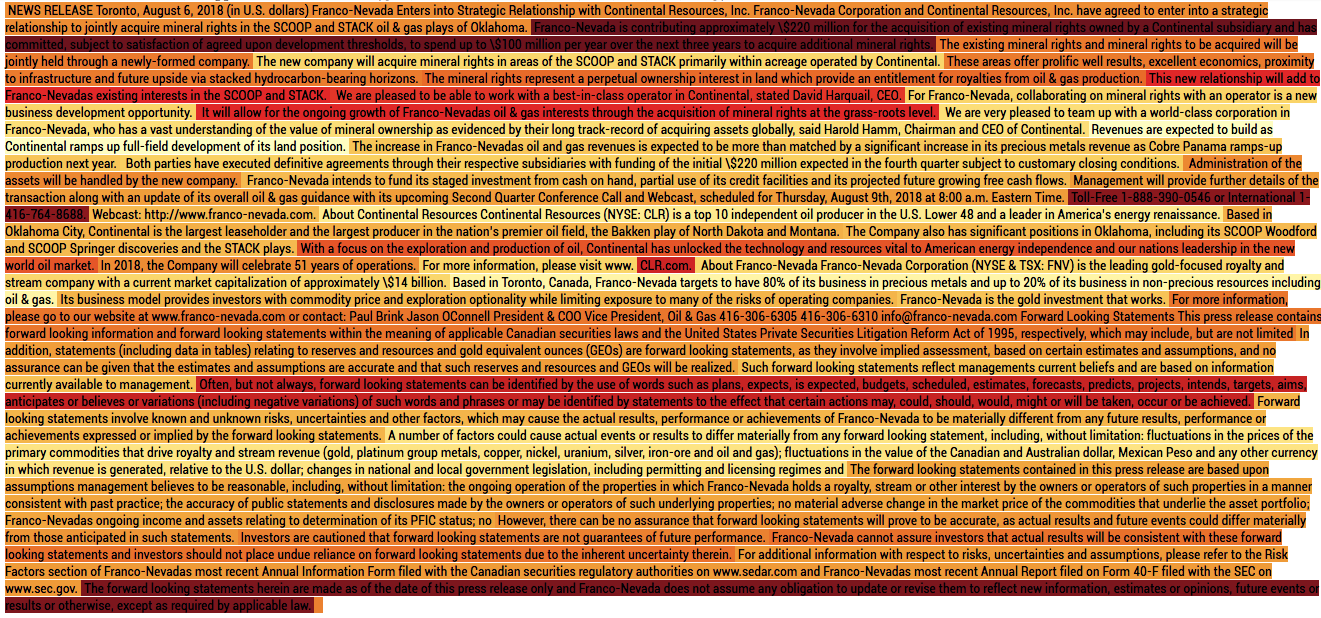}
    \end{subfigure}
            \caption{Examples of documents within most confident predictions of increased volumes with attentions focus based on attention vector norms. Darker color indicates stronger focus.}
        \label{fig:examples_un}
\end{figure}

\begin{figure}
    \begin{subfigure}{1\linewidth}
        \includegraphics[width=1\textwidth]{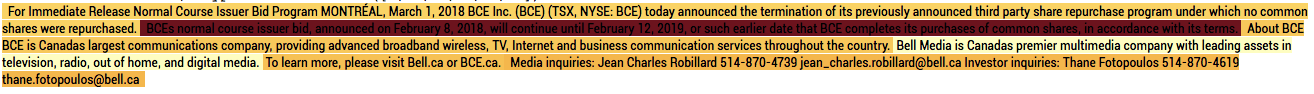}
    \end{subfigure}
        \begin{subfigure}{1\linewidth}
        \includegraphics[width=1\textwidth]{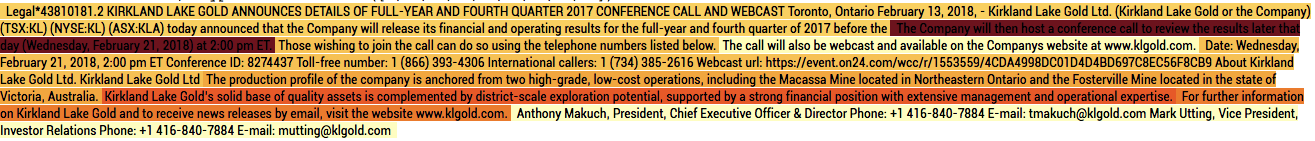}
    \end{subfigure}
            \caption{Examples of documents within most confident predictions of decreased volumes with attentions focus based on attention vector norms. Darker color indicates stronger focus.}
        \label{fig:examples_down}
\end{figure}

\newpage

\end{document}